\documentclass[conference]{IEEEtran}
\IEEEoverridecommandlockouts
\usepackage{cite}
\usepackage{amsmath,amssymb,amsfonts}
\usepackage{algorithmic}
\usepackage{graphicx}
\usepackage{textcomp}
\usepackage{xcolor}
\usepackage{booktabs,tabularx}
\usepackage{caption}

\graphicspath{{./figures/}}

\def\BibTeX{{\rm B\kern-.05em{\sc i\kern-.025em b}\kern-.08em
    T\kern-.1667em\lower.7ex\hbox{E}\kern-.125emX}}

\begin{document}

\title{Analyzing BEV Suitability and Charging Strategies Using Italian Driving Data}



\author{\IEEEauthorblockN{Homa Jamalof, Luca Vassio, Danilo Giordano, Marco Mellia}
\IEEEauthorblockA{\textit{Politecnico di Torino}\\
Turin, Italy \\
\{homa.jamalof, luca.vassio, danilo.giordano, marco.mellia\}@polito.it}
\and
\IEEEauthorblockN{Claudio De Tommasi}
\IEEEauthorblockA{\textit{UnipolTech SpA}\\
Turin, Italy \\
claudio.detommasi@unipoltech.it}
}

    \maketitle

\begin{abstract}
Battery Electric Vehicles (BEVs) are rapidly evolving from a niche alternative to an established option for private transportation, often replacing Internal Combustion Engine (ICE) vehicles. Despite growing interest, significant barriers remain, including range anxiety, the inconvenience associated with public charging stations, and higher costs.
This study analyses extensive telemetry data collected from 10,441 users using ICE vehicles in an Italian province to assess the potential for switching to BEVs without changing current travel behaviour. We evaluate to what extent the BEV models can fulfil their mobility needs under different charging scenarios. To do so, we replicate trips and parking events, simulating and monitoring the battery state of charge. 
The analysis reveals the compromises between charging behaviours and limited BEV autonomy. Assuming access to overnight charging, at least 35\% of the users could already adopt even low-capacity BEVs. 
\end{abstract}


\section{Introduction}

Transport is one of the main causes of anthropogenic greenhouse gas emissions. 
The widespread adoption of electric vehicles is widely seen as an effective strategy to reduce greenhouse gas emissions due to their higher energy efficiency compared to Internal Combustion Engine (ICE) vehicles.
Accordingly, governments and car manufacturers are investing heavily in EV technologies to achieve significant emission reductions \cite{KUMAR2025104367}. 
Pure Battery Electric Vehicles (BEVs) operate exclusively with stored electrical energy and therefore require an external power source to recharge their batteries \cite{app132312877}. BEVs can be charged at private or public stations, using different power levels. 
Although high‐power fast DC chargers offer significant convenience on long trips, 
slower AC chargers are more common and are suitable for overnight or daytime charging at home and work \cite{XU201768}.

Range anxiety, long recharge times, unavailable charging stations, and upfront purchase and energy costs slow down BEV adoption and its associated societal benefits. Advances in battery capacity and charging speed have yet to fully mitigate these concerns \cite{ZHANG2022110}. 
The primary objective of this study is to perform a multi‐scenario simulation of the transition from internal combustion engine (ICE) vehicles to BEV models, using real‐world driving data from an Italian province (Asti).
These data are supplied by UnipolTech, a telematics company providing mobility solutions for data-driven services and technologies for insurance services. The data comprises more than 10 thousand users with a car and 10 million trips over 1 year. 
By replicating each user's trips with different BEV models varying in battery capacity, charging power, and road-type performance, we evaluate which models are best suited to the different user routines.
In addition to different BEV models, we compare several charging behaviours, from users charging at night with slow AC chargers (e.g., a user with a wallbox) to users charging with fast DC chargers only when the battery State of Charge (SoC) is low. 

This multi‐scenario approach allows us to analyse how BEVs can replace ICE vehicles under different charging behaviours and travel patterns, highlighting the key factors for a successful transition to electric mobility.

Our results show that the current technology in terms of consumption, battery capacity, and charging power is suited for part of the population, but not yet for all. 
Indeed, even assuming the availability of charging opportunities and disregarding costs, many users cannot perform all of their ICE trips without running out of battery. With the most favorable charging policy and the largest battery vehicle, 72\% of the users can satisfy 100\% of their trips without changing their driving habits. 

\section{Related Works}

Numerous studies explored the practical feasibility of replacing internal-combustion vehicles with battery electric vehicles, each contributing valuable insights while also exhibiting limitations that our work seeks to address.

In an early empirical assessment, authors of \cite{GREAVES2014226} analysed five weeks of trip data for 166 privately-owned vehicles under six hypothetical battery-capacity scenarios (8–36 kWh) and slow AC charging powers (2.4–7.2 kW AC) to evaluate daily driving feasibility. While results showed that even small-battery BEVs could meet most trip demands, the reduced number of users limits their statistical validity. 
In \cite{Paffumi2015}, the authors leveraged one month of GPS traces in two Italian cities to simulate fourteen distinct recharging strategies across six BEV categories. Their algorithm checked the state of charge sufficiency on a trip-by-trip basis.
Both these works overlooked model-specific consumption profiles and temporal variability in driving patterns.   

Authors of \cite{9543680} advanced the scale and temporal span of such analyses by exploiting one year of telematics data for over 52,000 vehicles. They introduced the Daily Vehicle Kilometers Traveled (DVKT) metric and flagged Critical Days when DVKT exceeded a fixed 200\,km BEV range.
However, this approach assumed a uniform 200 km range available each morning and did not update the state of charge on a per-trip basis. 

More recently, the authors of \cite{GIOVANNICOLOMBO2024100782} examined 200 vehicles using trip records to investigate three charging powers and three BEV models. Although their work highlighted the potential effects of private-parking availability, it lacked variation in charging policies and employed a single consumption value. 
Finally, in \cite{su17093983}, the authors utilise a year of data for 226,000 vehicles in three cities to compute a functional compatibility index under a single 300 km-range BEV and single charging power. By applying density-based clustering to infer home-charging locations, they achieve a mobility-index assessment, yet fail to diversify BEV ranges and charging rates.

In summary, prior work has moved from short‐duration, small‐sample investigations to large‐scale, year‐long telematics studies, yet most still rely on average consumption rates, static state‐of‐charge assumptions, or limited temporal scopes. Our research advances this field by implementing a multi-scenario simulation framework that: (1) applies vehicle- and road-type-specific energy profiles, (2) updates state-of-charge on a per-trip basis, and (3) captures seasonal and spatial variations in charging behavior. This approach enables a more granular and realistic assessment of the transition from ICE vehicles to BEVs under diverse charging regimes.

\section{Dataset Details and Characterization}

We use data from telematics supplied by the UnipolTech company.
Data covers the period from 1 October 2023 to 30 September 2024 for 10,441  private users with a car registered in the province of Asti, Italy.\footnote{
All records were fully anonymised, without providing any personal details of vehicle owners and geolocalization, in accordance with European privacy regulations.}

Telematic boxes are permanently installed on each car and collect data for each trip, which corresponds to a cycle ignition-on to ignition-off. The data contains the vehicle anonymised ID, the start and end timestamps, and the covered distance across road categories (urban, extra-urban, highway).
In total, 13,064,904 trips within an ignition cycle were captured.

From the trip sequence, we obtain the parking events as the inter-time between the end of a trip and the start of its successor.  
To clean the dataset, we discard parking events with durations shorter than 2\,min, merging the surrounding trips. Additionally, we discarded any trips lasting under 1 minute or exceeding 12 hours, those covering less than 5 meters or over 800 kilometers, and trips with average speeds below 5 km/h or above 130 km/h. 
After cleaning, the dataset comprises 10,141,809 trips and more than 98,889,757 km driven distance. 
      






\begin{figure}[h!]
  \centering
  \includegraphics[width=0.95\linewidth]{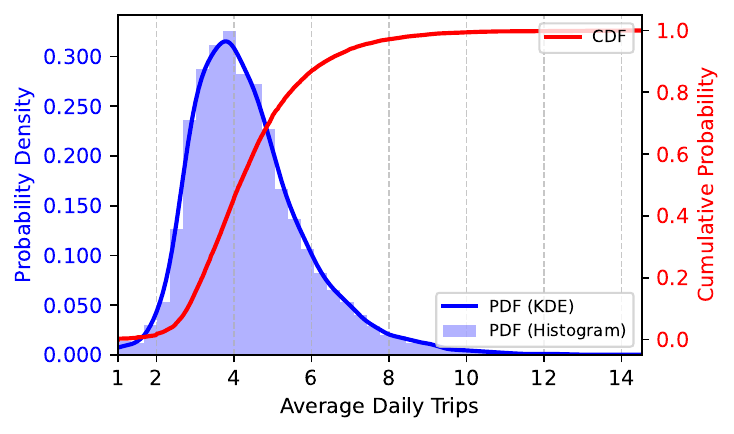}
  \caption{ PDF and CDF of average number of daily trips per user (active days). X-axis is limited to 99.9\% of the users.}
  \label{avg_daily_trips}
\end{figure}

\begin{figure}[h!]
  \centering
  \includegraphics[width=0.95\linewidth]{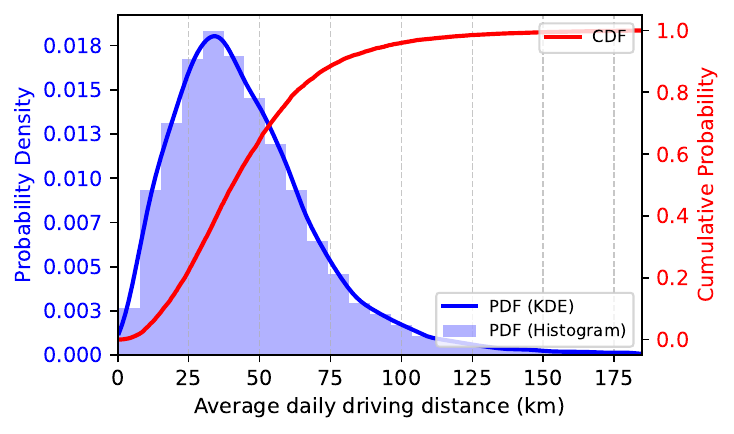}
  \caption{PDF and CDF of average daily driving distance (km) per user (active days). X-axis is limited to 99.9\% of the users.}
  \label{fig:avg_daily_distance}
\end{figure}

\begin{figure}[h!]
  \centering
  \includegraphics[width=0.95\linewidth]{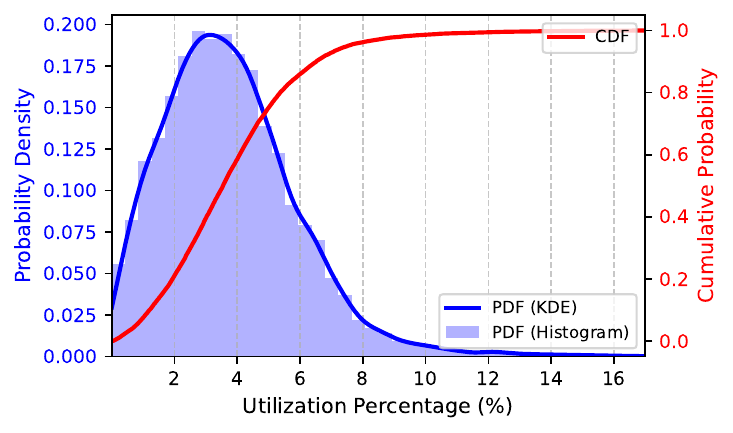}
  \caption{PDF and CDF of utilisation percentage per user on active days. X-axis is limited to 99.9\% of the users.}
  \label{fig:avg_daily_time}
\end{figure}

To characterize the behaviour of users, we report in Figure \ref{avg_daily_trips} the distribution of the average number of daily trips on active days (i.e., days with at least a trip). Half of the users perform fewer than 5 trips per day. Still, 10\% of the users have more than 8 daily trips. 
Figure \ref{fig:avg_daily_distance} reports the distribution of the average daily distance on the active days. On median, users drive an average of 40 km per active day, with only 4.5\%  exceeding 100 km. 
Finally, Figure \ref{fig:avg_daily_time} shows the distribution of the utilization of users on active days. This is computed on active days as the percentage of time the car is driven, i.e., the daily sum of trip durations divided by 24 hours. Notice how on more than 90\% of the active days, the car is driven less than 2 hours (corresponding to 8.3\% of utilization percentage).

These findings confirm that routine daily distances are easily covered by the autonomy of contemporary BEVs, including lower-capacity models (see Table \ref{tab:bev_specs}). 

\section{Simulating BEVs charge and discharge}

We exactly replicate the trips recorded by the ICE vehicles, without modifying their itineraries or schedules to accommodate battery constraints.
We consider multiple BEV model types, and then simulate their state of charge by updating it according to the travelled distance and consumption on the road type.


We updated the SoC when the car is parked and charged, considering four charging policies, inspired by the literature \cite{hardman2018review,ashkrof2020analysis,sun2021uncovering,andrenacci2023literature}:
\begin{enumerate}
 \item \textbf{Scenario 1.} From 08:00\,h to 20:00\,h on Monday through Friday, any continuous parking period of six hours or longer enables 7.4\,kW AC charging when $\mathrm{SoC}<75\%$. 
 This is a typical scenario for those users charging while at the workplace.
  \item \textbf{Scenario 2.}  Whenever $\mathrm{SoC}<25\%$, regardless of day or hour, AC charging at 7.4\,kW is applied during any parking period of six hours or more. This is a scenario where users charge only when the SoC is low and are attentive to costs.
  \item \textbf{Scenario 3.} The vehicle is charged via a 7.4\,kW AC connection when it is parked for at least six hours between 20:00\,h and 08:00\,h and its state of charge falls below 75\% ($\mathrm{SoC}<75\%$).
 This is a typical behaviour for those users with available overnight charging at home.
  \item \textbf{Scenario 4.} If $\mathrm{SoC}<25\%$,  at any time, a parking event of 20\,min or more triggers a 50\,kW DC fast-charge session. This is a scenario where users charge only when the SoC is low, and they want it to be fast, regardless of costs.  
 
\end{enumerate}
The scenarios encompass both slow AC charging, including residential and workplace setups, and fast DC charging, as typically deployed along highways. To limit the number of charges, we used two SoC thresholds (25\% and 75\%). 

\begin{table}[t]
\centering
\caption{Technical specifications of the four BEV models used in our trip‐replication simulations \cite{evdatabase}.}
\label{tab:bev_specs}
\scriptsize
\begin{tabular}{lccc}
\toprule
\textbf{} & \textbf{Usable net} & \textbf{Estimated} & \textbf{Consumption}\\
 \textbf{Vehicle model} & \textbf{capacity } & \textbf{autonomy} & \textbf{urban/highway/comb.} \\
 \textbf{} & \textbf{(kWh)} & \textbf{range (km)} & \textbf{(Wh/km)}\\
\midrule
Fiat    500e        & 21.3 & 135 & 101 / 170 / 133 \\
Renault  Megane E-Tech          & 40.0 & 260 & 103 / 167 / 133 \\
Tesla    Model 3          & 57.5 & 420 & ~93 / 142 / 116  \\
Audi     A6 e-tron        & 94.9 & 610 & 109 / 161 / 134 \\
\bottomrule
\end{tabular}
\end{table}

We consider four commercially available BEV car models to represent a spectrum of battery capacities and driving ranges.
Specifications for the four BEV models are summarized in Table \ref{tab:bev_specs}. 
We obtained the vehicle specifications and consumption figures from the EV Database website \cite{evdatabase}, where consumption values correspond to mild ambient conditions
and a constant speed of 110 km/h for highways.
Note that our uniquely detailed telematics dataset allows for a segmentation-based energy-use calculation. 

Simulating the trips and the corresponding discharge process, when the SoC reaches 0, we mark the corresponding trip as not feasible.
The SoC then remains 0 until a charging event is initiated. Any trip in between will be marked as not feasible.  

We focus on three metrics: (i) the percentage of feasible trips, (ii) the monthly number of charges, and (iii) the average SoC percentage after a trip. We compute these metrics for each user in the dataset and report statistics on the obtained distributions. 




\section{Results}

\begin{figure}
    \centering
    \includegraphics[width=\columnwidth]{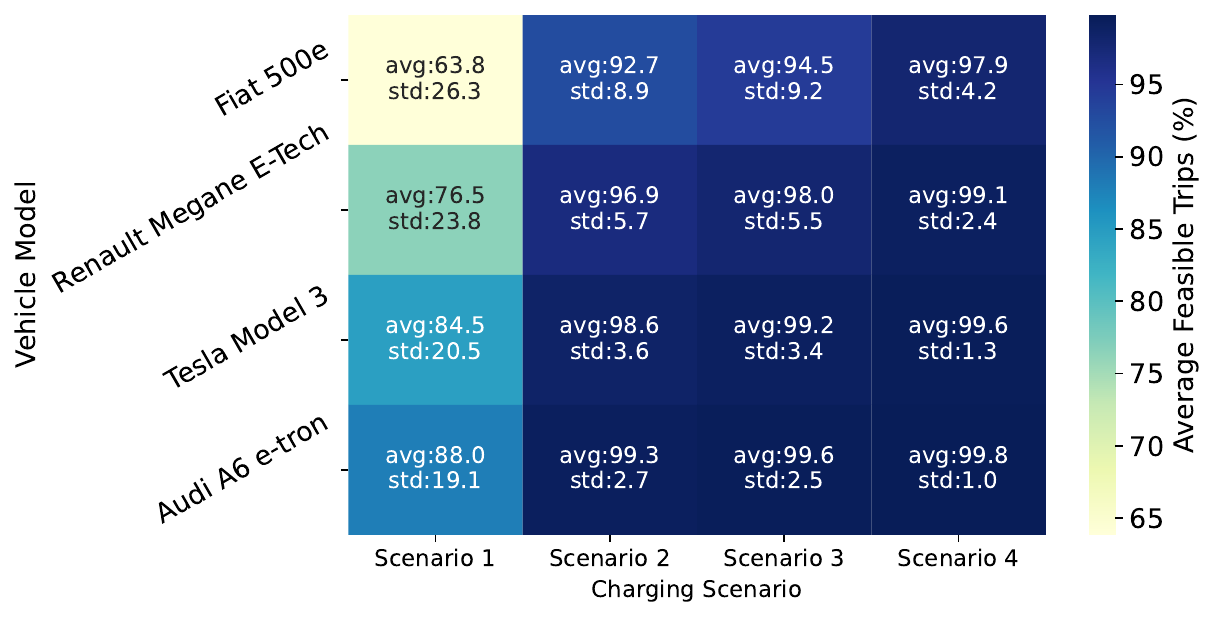}
    \caption{Feasible trip percentage by charging scenario and vehicle model. Color represents the average among users.
    }
    \label{fig:heatmap_feasible_trip}
\end{figure}


We evaluate the performance of the four proposed charging policies and the four vehicle models. 
Figure~\ref{fig:heatmap_feasible_trip} shows the average percentage of feasible trips across the different options. 

\begin{figure}
  \centering
  \includegraphics[width=0.95\columnwidth]{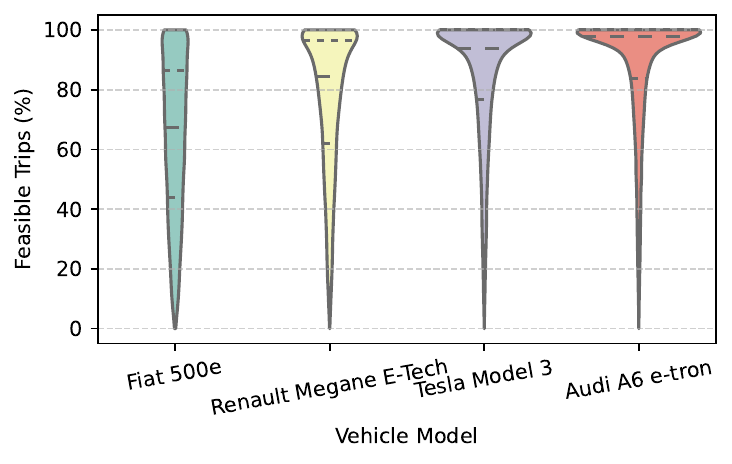}
\caption{Feasible trip percentage for charging scenario 1. Violin plots report the distribution among users, with marked quartiles. }
  \label{fig:feasible_scenario2}
\end{figure}

Scenario 1 yields the lowest performance. 
On average, users are able to complete only 64\% to 88\% of their trips under this scenario.
The distribution among users of feasible trip percentages for Scenario 1 is shown in Figure~\ref{fig:feasible_scenario2} using violin plots. 
If we define suitable users as those able to complete at least 99\% of their trips, a low-capacity BEV such as the Fiat 500e proves to be unsuitable for 93\% of the users.  
Performance improves with vehicles offering greater range. For the Audi A6 e-tron, which has the highest autonomy among the considered models, this policy is potentially viable for 43\% of the users. Nonetheless, this means that charging only during working hours and working days is not a feasible pattern for most users and entails substantial compromises. 

For other charging policies, differences in the average feasible trip percentage are less pronounced. Scenario 2 allows many users to complete nearly all of their trips, while Scenario 3 and Scenario 4 further improve the results.
For example, scenario 3 shows that overnight charging with the Fiat 500e is already suitable for 35\% of the users (with 92.7\% average feasible trips).  

\begin{figure}
  \centering
  \includegraphics[width=0.95\columnwidth]{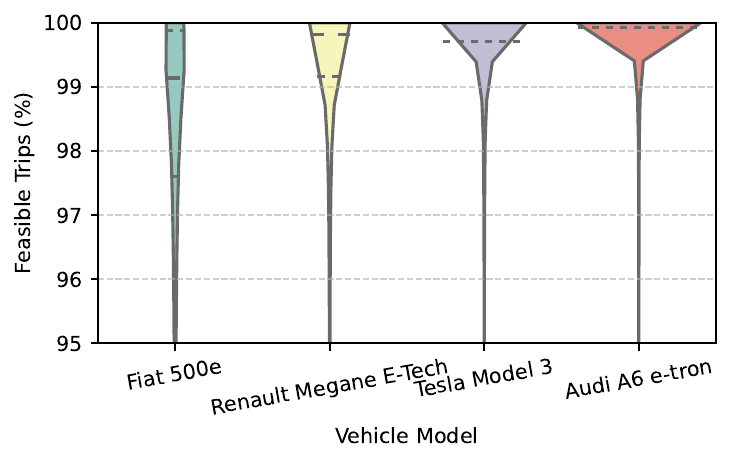}
\caption{Feasible trip percentage for charging scenario 4. Violin plots report the distribution among users, with marked quartiles. Notice that the y-axis is limited from 95\% to 100\% feasible trips percentage for readability.}
  \label{fig:feasible_scenario4}
\end{figure}

Scenario 4 performs best overall. Its results are reported in Figure \ref{fig:feasible_scenario4}.  The average feasible trip percentages exceed 98\% for all vehicle models except the Fiat 500e. In this case, the Fiat 500 appears suitable for 53\% of the users.
The best-case scenario is achieved with the largest battery vehicle (Audi A6 e-tron), where 95\% of users appear suitable. In this case, 72\% of the users can even satisfy 100\% of their trips. 


These findings indicate that, with the current capabilities of BEV technology, full electrification remains inadequate to accommodate the mobility needs of all users in the studied Italian province.
Still, many users are already suited for a BEV car, given that they have a convenient charging station at home or at the workplace, or have access to a fast one, or the possibility to use a large capacity BEV.  Notice that, in this work, we did not take into account costs, nor the availability of a charging station.

\begin{figure}
  \centering
  \includegraphics[width=0.95\columnwidth]{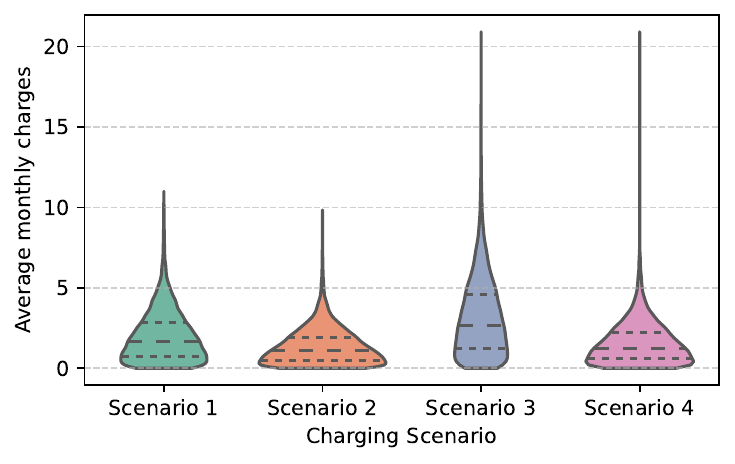}
\caption{Distribution among users of the average monthly number of charging events with the Audi A6 e-tron vehicle, changing charging scenario. 
}
  \label{fig:charges_audi}
\end{figure}

\begin{figure}[h!]
  \centering
  \includegraphics[width=0.95\linewidth]{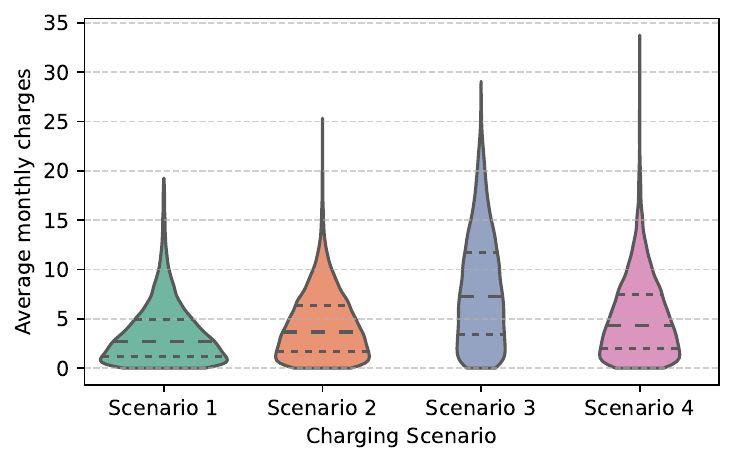}
\caption{Distribution among users of the average monthly number of charging events with the Fiat 500e vehicle, changing charging scenario.}
  \label{fig:charges_fiat}
\end{figure}

Then, we evaluate the number of charging sessions required under each policy. This is a critical metric, as users generally prefer to meet their mobility needs with minimal inconvenience related to locating charging stations and repeatedly plugging and unplugging the vehicle.

To illustrate this, we focus on the Audi A6 e-tron, the vehicle with the highest overall performance, achieving more than 99\% of feasible trips in Scenarios 2, 3, and 4. 
Figure~\ref{fig:charges_audi} shows the distribution of monthly charging sessions for this model. Although Scenarios 2, 3, and 4 provide similarly high suitability in terms of feasible trips, the number of charging sessions varies significantly. Notably, Scenarios 2 and 4 reduce the required number of charges by nearly half compared to Scenario 3. In Scenario 2, the median number of monthly charges is just 1.1, increasing to as many as 9.8 for users with long-distance driving demands.

We also report in Figure \ref{fig:charges_fiat} the average monthly number of charges for the Fiat 500e. Remember that, with this vehicle, the majority of users would not be able to satisfy their trips (less than 99\% feasible trips), under any charging policy. Still, users would need to charge their cars many times, with third-quartile users needing more than 12 times a month under Scenario 3. This is critical and could be considered unacceptable by many users, even if they accept the reduced trip feasibility performance.

\begin{figure}
    \centering
    \includegraphics[width=\columnwidth]{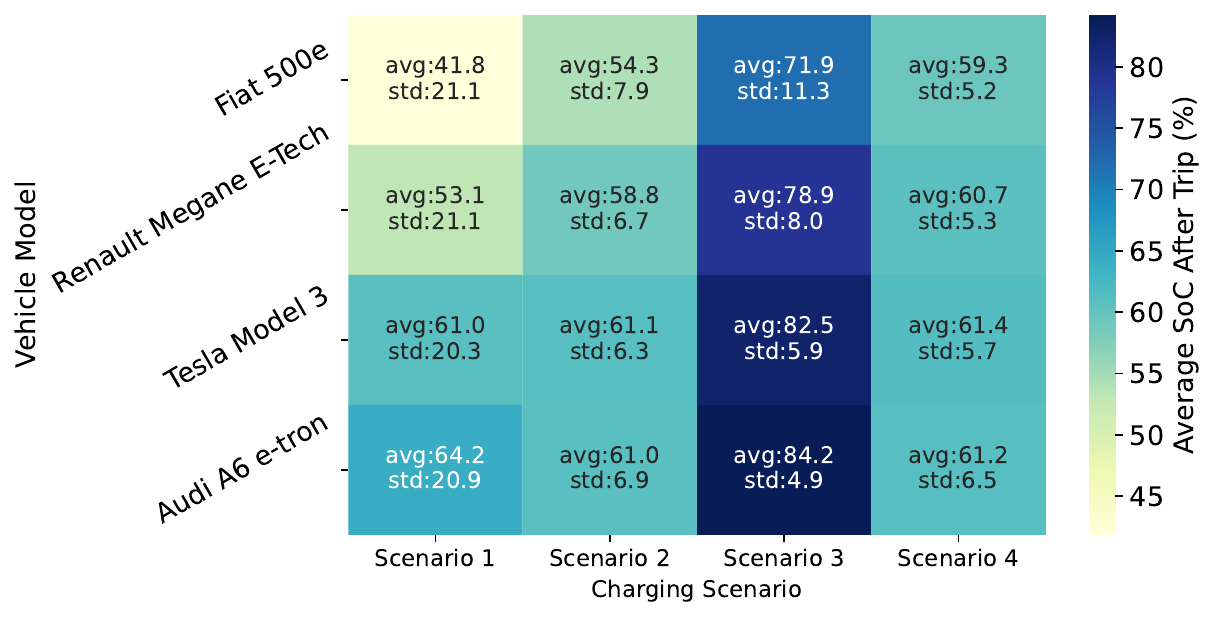}
    \caption{Average SoC percentage after trip by scenario and car model.}
    \label{fig:heatmap_avg_soc_after_trip}
\end{figure}

Finally, we report in Figure \ref{fig:heatmap_avg_soc_after_trip} the average SoC percentage after a trip across the different options of vehicle models and charging scenarios. This metric complements the one related to feasible trips. Indeed, a low SoC after a trip might induce a range anxiety feeling to the user and negative stress, and might trigger changes in behaviour, such as a desire to charge the car and look for a charging station. Therefore, we believe the SoC percentage after a trip should be maximized. Results show how Scenario 3 (overnight charging) obtains the best performance and guarantees, on average, a SoC above 70\%, irrespective of the car. This improvement in SoC level might counter-balance the slightly less feasible trips with respect to Scenario 4 (DC charging, but only when SoC goes below 25\%, see Figure \ref{fig:heatmap_feasible_trip}).

\section{Limitations and Conclusions}

We evaluated the transition feasibility from a large population of an Italian province, under varying charging regimes, travel demands, and vehicle models available on the market. 
We showed whether the travel patterns of part of the population are compatible with the current BEV models. Indeed, at least 35\% of the users could already adopt BEVs, assuming access to overnight charging (scenario 3, with Fiat 500e). Still, even the best-performing charging policy (scenario 4, with Audi A6 e-tron) and the largest battery vehicle are not enough for 5\% of the population. 

Several simplifying assumptions may affect our simulation. First, we treated energy consumption as invariant to external factors such as ambient temperature and assumed constant charging power throughout a session, despite evidence that charge rates typically decline once a battery’s state of charge exceeds about 80\% \cite{5978239}. Second, we presumed that charging infrastructure was accessible at every parking event, since precise user and charger locations were unavailable. Finally, we did not conduct a full cost–benefit analysis, omitting considerations of long‐term operational and maintenance expenses, as well as fluctuations in electricity prices, which would be necessary to evaluate the economics of different charging strategies.  
Finally, applying the methodology to other regions would validate generalizability and inform localized policy and infrastructure planning.

\section{Acknowledgment}
The authors thank UnipolTech (Unipol Group) for their support and for providing access to extensive vehicle telematics data, which formed the foundation of this research.

\bibliographystyle{IEEEtran}
\bibliography{biblio}

\end{document}